# A Three-Charge-Coupled-Device Camera Based on Multiplexed Volume Hologram

**Sun-Young Choi[1], Vladimir Pershin[2]**

[1]*School of Electrical Engineering, Soongsil University, Korea*

[2]*Department of Computer and Information Technology, Kharkiv Polytech Institute, Ukraine*

**Abstract:** We present a three-Charge-Coupled-Device (three-CCD) camera constructed upon multiplexed volume holographic gratings. Three incoherent gratings are written on the same holographic plate to guide the incident light into the three monochromatic channels, where images of red, green and blue light are recorded separately by CCD cameras. The reconstructed chromatic image from the monochromatic images shows superior qualities over the images taken by a single CCD camera.

**Key words:** three-CCD camera, volume hologram

# 1. Introduction

A three-CCD camera system uses three charge-coupled devices (CCD) to achieve superior imaging capability compared to single detector imaging systems .A special trichroic prism beam splitter is usually used for color separation to divide light into three primary colors: red, green, and blue which are directed to three separate CCDs. Final recombination of three images yields higher resolution and SNR compared to a single CCD imaging system [1-2]. In addition, the colors produced by a three-CCD camera are more vivid, lifelike, and accurate. As such, it is used widely in the fields of film production, astronomy, colormetry, as well as advanced surgical imaging [3-6]. Dividing the incoming beam into three constituent primary colors red (R), blue (B), and green (G) is the key to the functioning of a three-CCD system. In a conventional three-CCD camera, this is done by using a trichroic prism assembly which consists of a low-pass, a high-pass and a band-pass filters for directing the appropriate band of wavelength ranges to respective CCD's [7, 8]. It is crucial to design an appropriate wavelength filtering system for a high yield three-CCD camera system. In recent years, a great deal of research has been done about wavelength multiplexing thick holohraphic gratings [9-15]. Theoretical and experimental studies have shown that these volume gratings can be useful in a large number of practical applications such as devices for laser beam deflection, modulation, coupling, filtering, etc. [11-16]. In this paper, we present a three-CCD imaging system with optical color separation scheme utilizing a set of incoherently multiplexed volume holographic gratings. In this scheme, three incoherently multiplexed gratings on a holographic substrate diffract the incoming light into the three primary R, G and B monochromatic channels. The images obtained from these three channels are recombined to form the final output image. We have also measured the aberration of this imaging system. Images of a test object obtained using the proposed three-CCD camera system are compared with a single CCD camera imager. The comparison shows a higher image quality for our imaging scheme.

## 2. Writing the hologram

To construct a holographic director to divide the incoming light into three chromatic R, G, and B channels, we multiplexed three incoherent gratings on a single volume holographic plate with a 75 μm thick photo-sensitive storage medium. Figure-1 depicts the setup for writing such a hologram. A frequency-doubled Nd:YAG laser with a central wavelength of 532nm is used as the light source. After being divided by a beam spliter (BS), the reflected and transmitted components are coupled into two separated arms consisting of a spatial filter (SF) and a lens (L) to produce high quality collimated light. These beams are combined at the holographic plate (HP) to record the fringe pattern. Three holograms are recorded at $0^o$, $120^o$, and $240^o$ rotations of the HP, which is shown in Figure-2. The effect of one grating over the diffraction efficiency of the others is assumed negligible in this paper.

## 3. Properties of the hologram

Diffraction of a plane wave at a sinusoidally stratified holographic grating has been investigated theoretically [9, 15-17]. The resultant diffraction is similar to the case of an acoustic grating. Among a number of theoretical treatments available on the ultrasonic-light diffraction phenomenon, one of the simple but successful approaches is the coupled-wave theory (CWT) [18]. Following the steps of CWT, one can calculate the diffraction efficiency of a thick hologram. This treatment covers several types of holographic gratings: transmission and reflection gratings, slanted and unslanted gratings with respect to the surface, and gratings consisting of a spatially modulated refractive index or the absorption coefficient or both.

Figure-3 shows the diagram of a thick holographic grating with slanted fringes with respect to the norm of the surface at angle $\varphi$. The grating vector, determined by the writing beams, is denoted as $\beta$. Light incident at angle $\theta$ to the norm of the surface has a wave number $k_1$ in the medium. The interaction of the light and the grating yields two outputs: A beam ($E_1$) that follows the same wave vector ($k_1$) as the input, a beam ($E_2$) diffracted by the grating with wave vector $k_2 = k_1 - \beta$. The diffraction efficiency of the grating is defined as $\eta = \left|\dfrac{E_2}{E_0}\right|^2$. From the CWT, the Bragg selectivity condition, under which maximum diffraction efficiency can be achieved, is $|k_2| = |k_1|$. For a given grating, this is equivalent to

$$\cos(\varphi - \theta) = \beta / 2k_1 \tag{1}$$

The Bragg condition can be violated by change in angle $\theta$ or wave number $k_1$. Under such circumstances, the diffraction efficiency of the grating is

$$\eta = \sin^2(\nu^2 + \xi^2)^{\frac{1}{2}} / (1 + \xi^2 / \nu^2) \tag{2}$$

where  $\nu = \kappa d/(c_R c_S)^{\frac{1}{2}}, \xi = -\frac{1}{2}d\frac{\vartheta}{c_S}, c_R = \cos\theta, c_s = \cos\theta - \frac{\beta}{k_1}\cos\varphi$

$\kappa$ called the coupling constant and $\vartheta$ the dephasing measurement, which are defined as

$$\kappa = \pi n / \lambda \tag{3}$$

$$\vartheta = \Delta\theta \cdot \beta \cdot \sin(\varphi - \theta) - \Delta\lambda \cdot \beta^2 / 4\pi n \tag{4}$$

Variation of the typical diffraction efficiency for a holographic gratings as a function of the incident angle and wavelength from the Bragg condition is shown in Figure-4.

## 4. Constructing the three-CCD camera

To construct the imaging system, we used the experimental setup shown in figure-5. A lens was placed in front of the HP. Three band-pass filters are placed after the HP to select particular wavelength (410nm, 532nm or 633nm) in the diffracted light. The transmission in each channel is recorded by CCD's behind the filters on the image plane. The monochromatic images produced by the CCD's are recombined to form the chromatic image of the test object by computer program.

## 5. Aberrations

Aberrations are critical criterion in evaluating imaging systems [19]. We measured the spherical aberration of our holographic Three-CCD imaging system with the setup shown in Figure-6. The aberration plots are shown in Figure-7. The purple line is the aberration of the Nikon Nikkor lenses used in Figure-5. The blue and yellow lines denote the aberration of the whole three-CCD imaging system in the y- and x-directions respectively. The system has a larger aberration in the y-direction as the ratio of $\frac{P_y}{P_x}$ gets bigger, where $P_y$ and $P_x$ denote the position of the object respectively in the x and y axes with respect to the center of the lens. Per our results, the hologram is spherical-aberration-free along the entire x-direction and in 60% of the y-direction. All the measurements were performed with fully open lens apertures. By adjusting the lens apertures, we can operate the system in the area without hologram aberration.

**6. Imaging with the three-CCD camera**

We test the imaging system described in section 3 with a canister illuminated by a Halogen light. Figure-8 (a) through (d) shows the R, G, B images taken by the CCD cameras and the reconstructed chromatic image. Figure-9 shows an image of the same test object using a single CCD imager. A comparison of figure-8(d) with figure-9 shows an outstanding ability of our system to image the object with high resolution. We did observe some vignetting effect which arises due to lens aperture size and the angular bandwidth of the hologram. This may be the major limitations of system's field of view (FoV). To eliminate the restriction, many measures can be taken such as increasing the angular bandwidth by using a thinner sample, improving the hologram's diffraction with new material selection [20] and optimizing the diffraction efficiency of the hologram by adjusting intensity of the writing beams.

**Summary**

We have designed a simple three-CCD imaging system based on multiplexed volume holographic gratings. The aberration of the system mostly comes from the imaging lenses. The images taken by the three-CCD system shows a superior image quality over the image taken by a single CCD camera, and has potential for use in high resolution imaging applications.

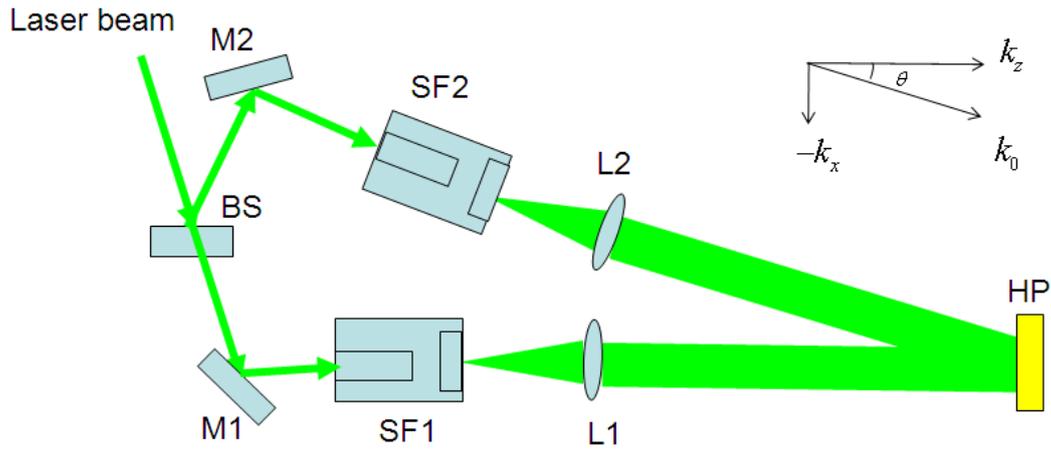

Figure-1 schematic diagram of writing the holographic gratings.

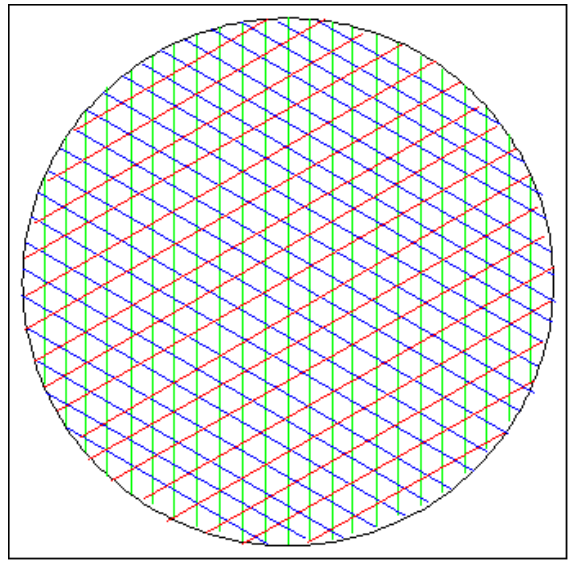

Figure-2 Diagram of the multiplexed holographic gratings for three primary colors

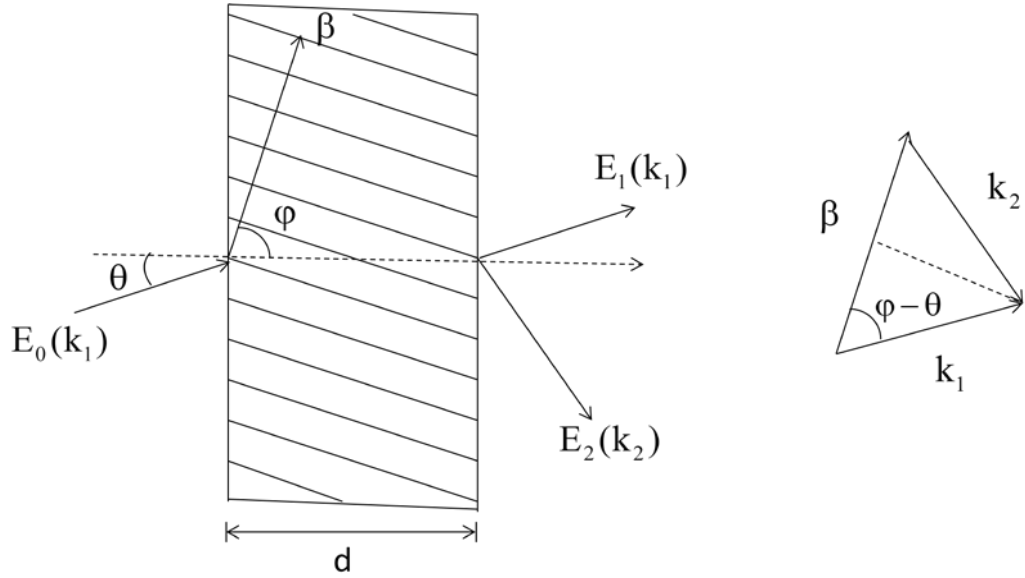

Figure-3 Block diagram of light diffracted by a thick hologram grating with slanted fringes.

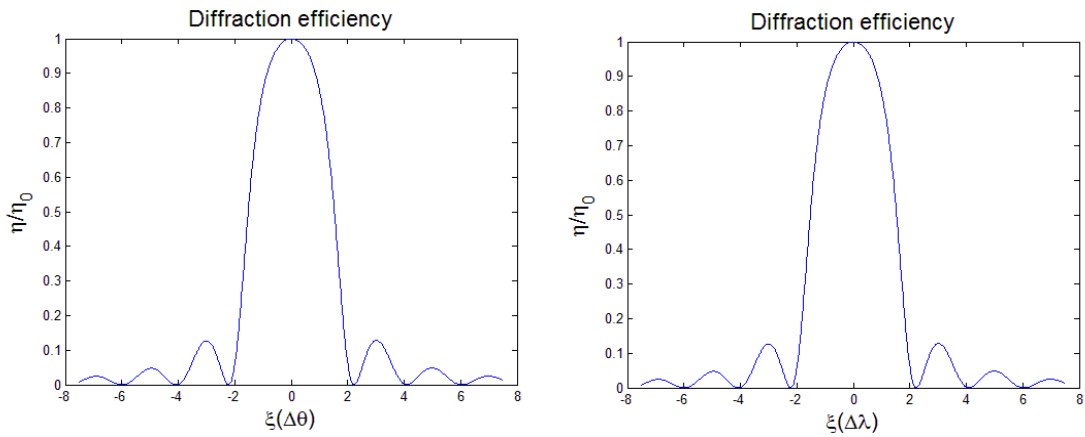

Figure-4 Diffraction efficiency for incident angle deviated from the Bragg angle.

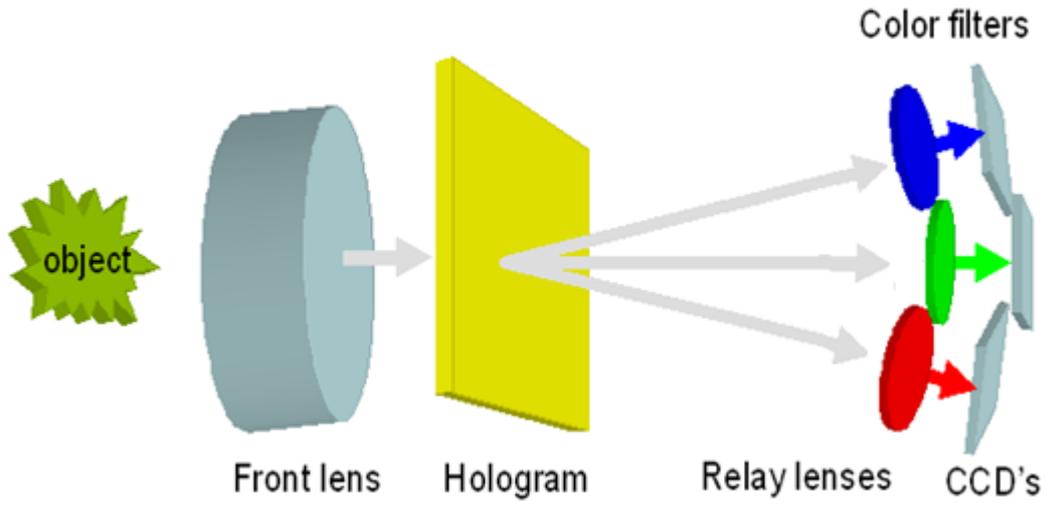

Figure-5 The 3 CCD imaging system based on multiplexed volume gratings.

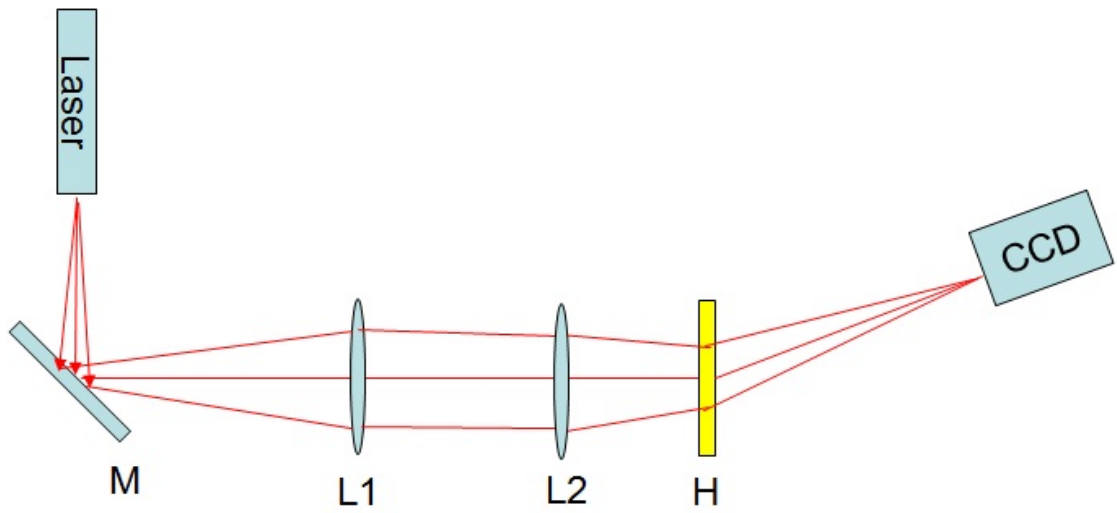

Figure-6 Setup for measuring spherical aberration of the 3 CCD imaging system.

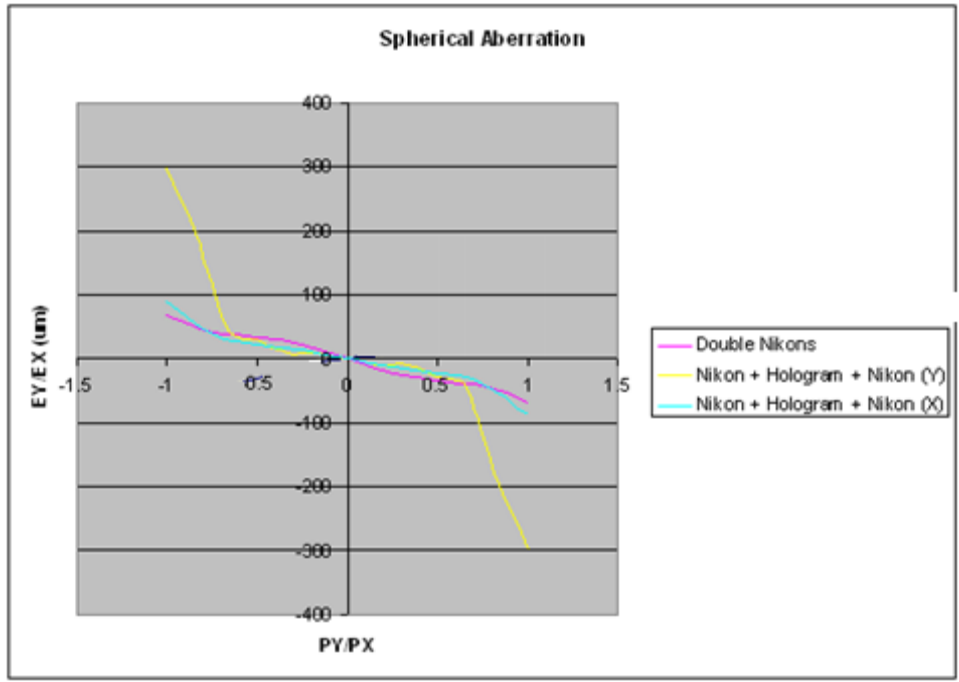

Figure-7 Spherical aberration plot of single lenses and the whole system

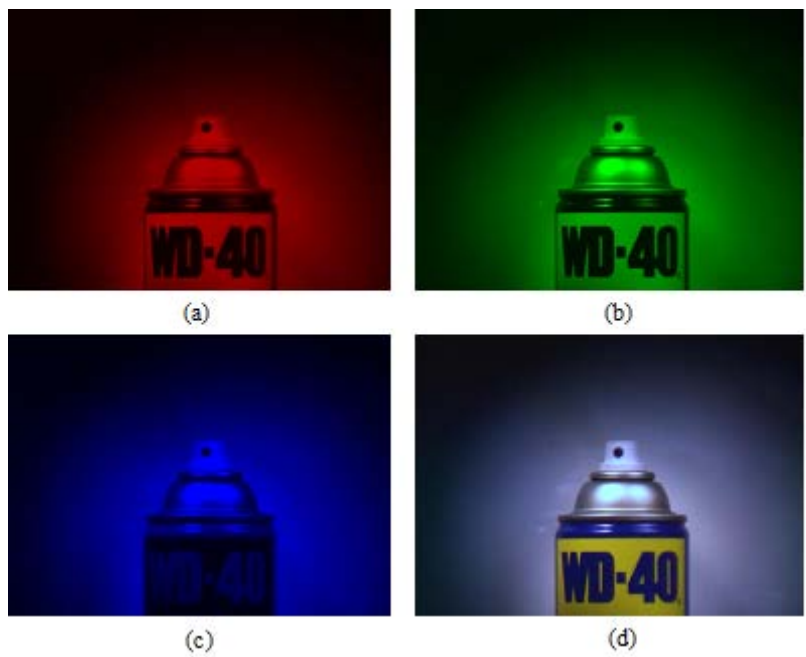

Figure-8 monochromatic and composite images of a canister

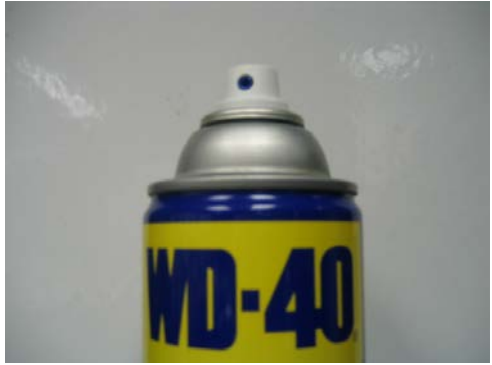

Figure-9 Canister image taken by a single CCD camera